\documentclass[12pt,superscriptaddress,reprint]{revtex4-1}
\usepackage{setspace}
\usepackage{amsmath}
\usepackage{relsize}
\usepackage{amssymb}
\usepackage{wasysym}
\usepackage{soul}
\usepackage{chemformula}
\usepackage{bm}
\usepackage{float}
\usepackage{graphicx}
\usepackage{epstopdf}
\usepackage[caption=false]{subfig}
\usepackage{stackengine}
\usepackage{verbatim}
\usepackage{makecell}
\usepackage{color, colortbl}
\usepackage{url}
\usepackage{etoolbox}
\usepackage{hyperref}
\graphicspath{{figures/}}
\patchcmd{\section}
{\centering}
{\raggedright}
{}
{}
\patchcmd{\subsection}
{\centering}
{\raggedright}
{}
{}

\begin{document}
\title{Anomalous Diffusion along Metal/Ceramic Interfaces}
\author{Aakash Kumar$^\dagger$}
\thanks{These authors contributed equally to this work.}
%\email[corresponding author:]{aakashk@seas.upenn.edu} 
\affiliation
{
	\mbox{Department of Materials Science and Engineering, University of Pennsylvania, Philadelphia PA 19104, USA}\\
}
\author{Hagit Barda}
\thanks{These authors contributed equally to this work.}
\affiliation
{
	\mbox{Department of Materials Science and Engineering, Technion - Israel Institute of Technology, 3200003 Haifa, Israel}\\
}

\author{Leonid Klinger}
\affiliation
{
	\mbox{Department of Materials Science and Engineering, Technion - Israel Institute of Technology, 3200003 Haifa, Israel}\\
}

\author{Michael W. Finnis}
\affiliation
{
	\mbox{Department of Materials and Department of Physics, Imperial College, London SW7 2AZ, UK}\\
}
\affiliation
{
    \mbox{Thomas Young Center, London SW7 2AZ, UK}\\
}

\author{Vincenzo Lordi}
\affiliation
{
	\mbox{Materials Science Division, Lawrence Livermore National Laboratory, Livermore, CA 94550, USA}
}

\author{Eugen Rabkin}
\affiliation
{
	\mbox{Department of Materials Science and Engineering, Technion - Israel Institute of Technology, 3200003 Haifa, Israel}\\
}

\author{David J. Srolovitz}
\email[corresponding author:]{ aakashk@seas.upenn.edu} 
\email[\\corresponding author:]{ srol@seas.upenn.edu}
\affiliation
{
	\mbox{Department of Materials Science and Engineering, University of Pennsylvania, Philadelphia PA 19104, USA}\\
}
\affiliation
{   
    \mbox{Department of Mechanical Engineering and Applied Mechanics, University of Pennsylvania, Philadelphia, PA 19104, USA}
}

%\date{\today}
\begin{abstract}
Hole formation in a polycrystalline Ni film on an $\alpha$-Al$_2$O$_3$ substrate coupled with a continuum diffusion analysis demonstrates that Ni diffusion along the Ni/$\alpha$-Al$_2$O$_3$  interface is surprisingly fast. \textit{Ab initio} calculations demonstrate that both Ni vacancy formation and migration energies at the coherent Ni/$\alpha$-Al$_2$O$_3$ interface are much smaller than in bulk Ni, suggesting that the activation energy for  diffusion along  coherent Ni/$\alpha$-Al$_2$O$_3$ interfaces is comparable to that along (incoherent/high angle) grain boundaries.  Based on these results, we develop a simple model for diffusion along metal/ceramic interfaces, apply it to a wide range of metal/ceramic systems and validate it with several  \textit{ab initio} calculations. These results suggest that fast metal diffusion along metal/ceramic interfaces should be  common, but is not universal. 
\end{abstract}

\maketitle
\noindent
Metal-ceramic interfaces are ubiquitous building blocks for a wide range of technologies, from semiconductor devices (metal/gate-oxides)~\cite{sah1991fundamentals} to thermal barrier coatings in gas-turbines (metal/Yttria Stabilized Zirconia (YSZ)~\cite{reed2008superalloys} to all-solid-state batteries (Li anode/electrolyte)~\cite{kraytsberg_review_2011}. 
Device performance thus directly depends on the integrity of these metal-ceramic interfaces. 
For example, in the case of Li-air batteries, ceramic coatings have been proposed (e.g., sapphire)
to prevent the degradation of the Li anode (prone to dendrite formation)~\cite{brandt_historical_1994,kazyak_improved_2015}. 
In CIGS-based (CuIn$_{(1-\mathrm{x})}$Ga$_\mathrm{x}$Se$_2$) photovoltaic cells, MoSe$_2$ forms at the Mo/CIGS interface~\cite{klinkert_new_2016}, degrading cell performance.  
In all of these examples, performance is affected by the transport of atoms along metal/ceramic interfaces; hence, rational device design demands improved understanding of transport along metal/ceramic interfaces~\cite{balluffi1982basic}. 
While an extensive literature exists on the mechanical strength~\cite{scheu2006interface}, atomic structure~\cite{ikuhara_high_1998}, and chemical composition~\cite{shashkov_atomic_1995} of metal-ceramic interfaces,  little is known about atomic diffusion along this channel. 
Diffusion along extended crystal defects (e.g., surfaces, dislocations, and grain boundaries (GBs)) is commonly much more rapid than bulk diffusion~\cite{barnes_diffusion_1950,garbrecht_dislocation-pipe_2017}. 
The widely-quoted hierarchy of diffusivities is \textit{D}$_\text{bulk}$ $\le$ \textit{D}$_\text{dislocation}$ $\le$ \textit{D}$_\text{GB}$ $\le$ \textit{D}$_\text{surface}$~\cite{n.a._gjostein_short_1973,mehrer2007diffusion}. 
Here, we focus on where metal/ceramic interfaces fall within this hierarchy, to discover the features that control such diffusion, and use these to predict its magnitude.
\begin{figure*}[!htb]
	\centering
	\includegraphics[width=1\linewidth]{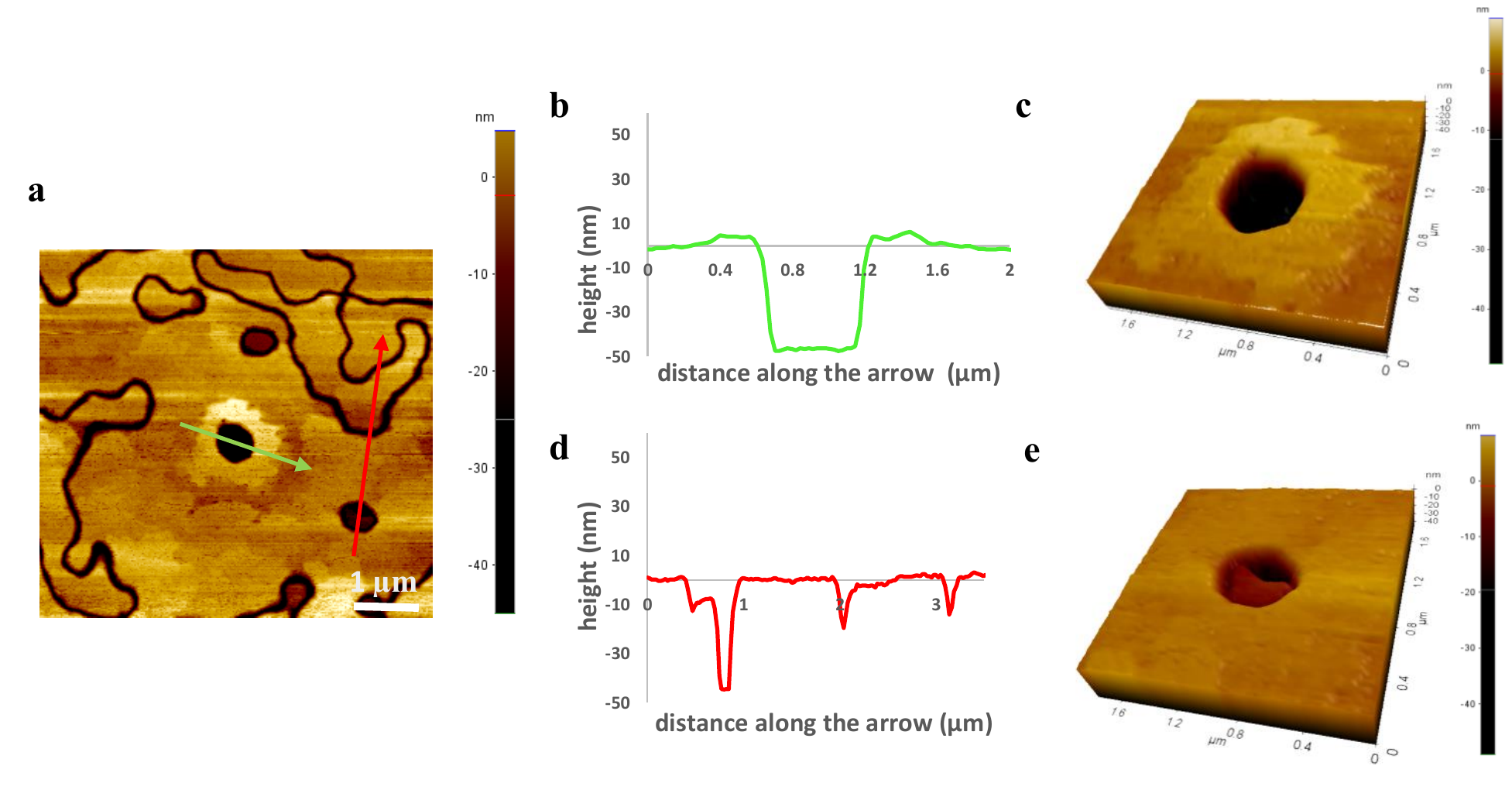}
	\caption{\textbf{Experimental observations of an annealed Ni film on sapphire.} \textbf{a},  Atomic Force Microscopy (AFM) topography image of a Ni film on sapphire after heat treatment, showing sunken grains with (green arrow) and without (red arrow) an elevated ridge around the edges. \textbf{b}, A linear topographic profile across the hole indicated using a green arrow in \textbf{a} showing a Ni ridge at the perimeter of the sunken grain. \textbf{c}, 3-D image of the same hole showing an elevated ridge surrounding the sunk grain. \textbf{d}, A linear topographic profile across the hole indicated using the red arrow in \textbf{a} showing no measurable ridge formation. \textbf{e},  3-D image around the same hole showing a hole left by grain sinking.}
	\label{fig:grain_sink}
\end{figure*}

There is indirect evidence to suggest that metal/ceramic interfaces may be high-diffusivity paths for metal atoms. 
For example, Gan \textit{et al.}~\cite{gan_effect_2006} showed that Cu diffusion along the (SiN,SiC)/Cu interface is faster than  in polycrystalline Cu.   
Arnaud \textit{et al.}~\cite{arnaud_evidence_2002} demonstrated that interface diffusion in Cu  interconnects (Cu on SiN) is faster than along Cu GBs. Recently,  indirect evidence of fast diffusion along metal/ceramic interfaces was found in the solid-state dewetting of metal films on oxide substrates~\cite{kovalenko_solid-state_2013,kosinova2015mechanisms}.  
It is reasonable to conjecture that diffusion along metal/ceramic interfaces will be comparable to that along other internal interfaces (e.g. GBs in metals). 
Yet, metal--oxygen bonds (in oxide ceramics) may be much stronger than those between metal atoms, suggesting rather that diffusion along metal/ceramic interfaces may be suppressed (relative to GBs in metals).  
Fast diffusion along GBs is usually attributed to their lower density or higher free volume (relative to grain interiors); a concept supported by the observation that diffusion along coherent twin boundaries appears no faster than in the bulk~\cite{minkwitz1999inclination}. 
In fact, Chen \textit{et al.}~\cite{chen_observation_2008}  showed that coherent twin boundaries  slowed electromigration in Cu. 
Nonetheless, there is evidence that diffusion along coherent (and semi-coherent) metal-ceramic interfaces is rapid~\cite{hieke2017annealing,hieke2017microstructural}.
This seems contrary to the association of fast transport with  low atomic density.
%\vspace*{-0.1cm}

In this work, we report experimental evidence of fast diffusion along the Ni/$\alpha$-Al$_2$O$_3$ (sapphire) interface. 
We explore the origin of this effect as a function of environment (O$_2$ partial pressure) using \textit{ab initio} calculations of point defects within these materials and near their interface. We establish that in this case, the dominant defects for interface transport are Ni vacancies and that the diffusivity of Ni along the interface is anomalously fast.  We then present a simple model that generalizes these results to a very wide range of metal/ceramic interface systems and validate the model through examination of additional metal/ceramic interface systems.
\vspace{-0.5cm}
\section*{E\lowercase{xperimental evidence of fast interface diffusion}}
\vspace{-0.4cm}
\noindent
We examine the case of an annealed 45 nm thick Ni film deposited on the (0001) surface of sapphire. 
All Ni grains have  $\left<111\right>$ surface normals with two in-plane orientations (rotated by 60$^{\circ}$ about the normal with respect to one another); this is referred to as a \emph{maze} microstructure.
We observed that the grain surfaces are flat, except for the presence of ridges and/or grooves at some GBs and holes within some of the grains~\cite{amram2014grain} (see Fig.~\ref{fig:grain_sink}a). 

The presence of isolated holes, not connected to GBs, is a clear indication that isolated/embedded grains (see Fig.~\ref{fig:grain_sink}a) sink and disappear, leaving through-thickness holes. 
These holes form in an early stage of solid-state dewetting. 
The hole indicated by the green arrow in Fig.~\ref{fig:grain_sink}a is surrounded by a slightly elevated ridge (see Fig.~\ref{fig:grain_sink}b,c). 
Integrating the profile around this hole from the AFM topography reveals that the volume contained in the ridge is $\sim0.5 \times 10^{-2}$ $\mu$m$^3$, while the volume of the material removed to form the hole is $\sim2$X as large. % ($\sim1 \times 10^{-2}$ $\mu$m$^3$). 
The ``sunken'' grain  indicated by the red arrow in Fig.~\ref{fig:grain_sink}a shows no such ridges (Fig.~\ref{fig:grain_sink}d,e). 
In the classic theory of GB grooving ~\cite{mullins1957theory}, all of the material removed around the GB goes into the ridge. 
Where did the missing Ni atoms go? How did these holes form?
\begin{figure*}[!htb]
	\centering
	\includegraphics[width=1\linewidth, scale=0.75]{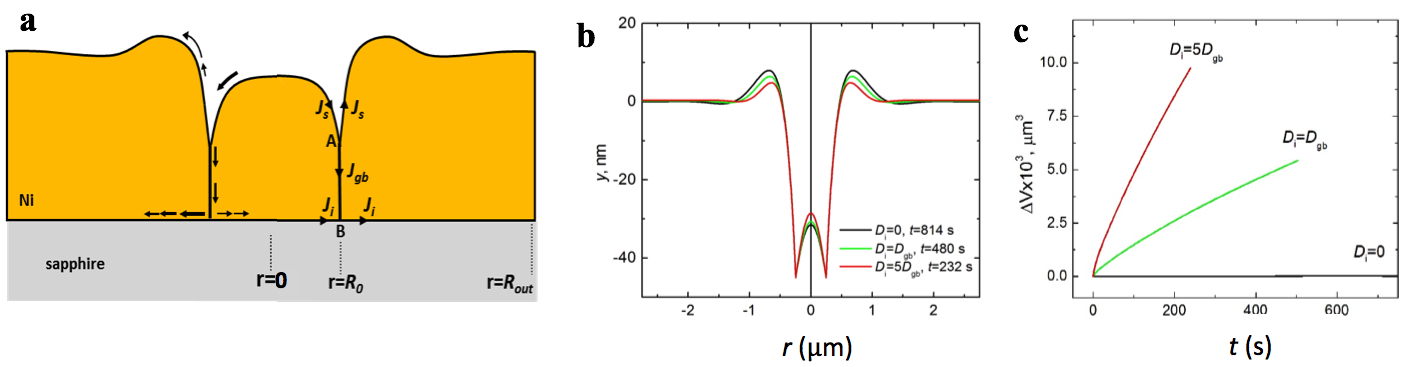}
	\caption{\textbf{Development and results of the morphology evolution model.} 
	\textbf{a},  The Ni/sapphire model showing Ni flux along the Ni surface ($J_s$), along the GB (A is the GB root position) between the sinking middle grain and the outer grain ($J_{gb}$), and along the Ni/sapphire interface ($J_i$) (B is the location of the intersection of the GB and the interface).  
	\textbf{b}, Surface topography profiles \textit{y(r)} for three different  interface diffusivities at the time $t$ the GB root hits the metal/ceramic interface at 700$^\circ$C. 
	\textbf{c}, Calculated ``missing'' Ni  volume $\Delta V$ vs. time. 
	}
	\label{fig:diffusion_plot}
\end{figure*}

We suggest that the missing Ni in the grooving/hole formation, diffuses down the GB and then along the Ni/sapphire interface (see Fig.~\ref{fig:diffusion_plot}a). 
For this process to continue, Ni diffusion  along the Ni/sapphire interface must be rapid. 
In this case, the film surrounding the sunken grain must thicken by the accretion of Ni atoms at the Ni/sapphire interface~\cite{amram2014grain}. 
Similar homogeneous thickening of a metal film associated with material redistribution along the metal-ceramic interface was recently reported in Al/sapphire~\cite{hieke2017annealing}. 

We test this hypothesis by developing a short-circuit diffusion model that describes the surface topography evolution of a thin metal film on a ceramic substrate via simultaneous surface, GB, and interface metal diffusion (see Fig.~\ref{fig:diffusion_plot}a  and Supplementary Information,~\hyperlink{linkSI}{SI} for details). 
Numerical solution of the evolution of the surface profile for the case of a small, axisymmetric grain of radius $R_0$ embedded in a continuous film is shown in Fig.~\ref{fig:diffusion_plot}b for three different interface diffusivities at the time when the GB groove hits the substrate (i.e., hole nucleation and onset of solid state dewetting).
This figure clearly shows that interface diffusion greatly enhances the rate of hole formation and reduces the amplitude of the elevated rim demonstrating that our model is consistent with the observations.

We analyze the experimental case of Fig.~\ref{fig:grain_sink}b,c (where a small ridge forms around the sinking grain) employing reasonable values of GB and surface diffusivities for Ni  via our model (see \hyperlink{linkSI}{SI}) in order to determine the interface diffusivity required to explain the ``missing'' Ni (the difference between the Ni forming the ridge and that from the sinking grain,
$\Delta V= 0.5 \times 10^{-2}$ $\mu$m$^3$ after a 10 minute annealing at 700$^{\circ}$C). These experimental observations yield $D_i \approx D_{gb}$ (for a random large angle GB in Ni~\cite{divinski2010grain}).  We note that given the variability of GB and surface diffusivity with bicrystallography and the uncertainty in the experimental measurements, we consider this to be an ``order of magnitude'' estimate.

This interface diffusivity is surprisingly large given that this metal/ceramic interface is nearly coherent and the strong bonding between this metal and ceramic.  We now address the question, ``why is this interface diffusivity so large?''  and, ``is this a generic finding for all metal/ceramic interfaces?''
\vspace{-0.5cm}
%\begin{flushleft}
\section*{F\lowercase{irst-principles modeling of point defects}}
%\end{flushleft}
\vspace{-0.4cm}
\noindent
To understand  fast transport along the Ni/sapphire interface, we first focus on point defect formation energetics using density functional theory (DFT) (to account for the complex bonding at the interface). 
Transmission electron microscopy examinations of the Ni/sapphire interface established two distinct orientation relationships. These are M1: Ni(111)[$1\bar{1}0$]$\|$Al$_2$O$_3$(0001)[11$\bar{2}$0] and M2: Ni(111)[$1\bar{2}1$]$\|$Al$_2$O$_3$(0001)[11$\bar{2}$0]~\cite{fogarassy2013growth,meltzman2012solid} (which is rotated by 30$^{\circ}$ about the surface normal from M1)~\cite{meltzman2012solid}. 
\begin{figure}[!htb]
	\centering
	\includegraphics[width=\linewidth]{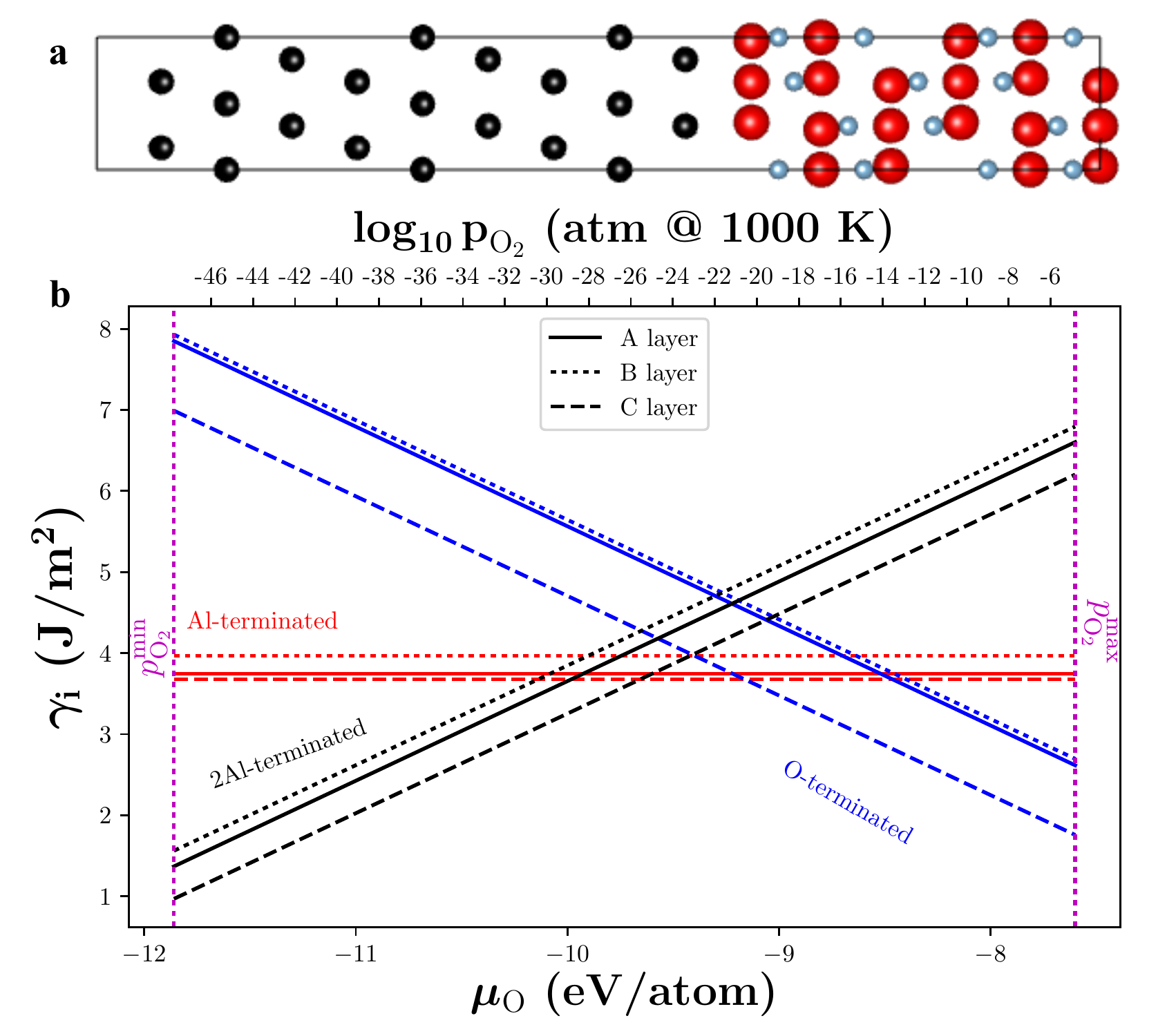}%
	\caption{\textbf{DFT prediction of the M1 Ni(111)[1$\bar{1}$0]$\|$Al$_2$O$_3$(0001)[11$\bar{2}$0] interface energy.} \textbf{a}, Unrelaxed, atomistic model of the (C-terminated) Ni/(O-terminated) sapphire interface. Ni, Al and O are shown in black, blue and red. 
	\textbf{b}, Interface energy of the Ni/sapphire interface considering the three Ni translations and the three sapphire terminations.
	}
	\label{fig:sapphire_unit_cell}
\end{figure}
Perfect interface coherency demands that Ni is biaxially strained -2.8$\%$ for M1 and +12.3$\%$ for M2. 
We focus on the less strained M1 interface (the M2 interface will have a much higher interface energy associated with a high density of misfit dislocations). 
The O-terminated Al$_2$O$_3$/Ni interface is shown in Fig.~\ref{fig:sapphire_unit_cell}a;  the  Al$_2$O$_3$ may also be terminated by one or two Al atom planes.
Examination of Fig.~\ref{fig:sapphire_unit_cell}b shows that the 2Al-terminated and O-terminated interfaces are stable over a wide range of oxygen chemical potentials; at high (low) $p_{\text{O}_2}$  the O (2Al )-terminated Ni/Al$_2$O$_3$ interfaces will be thermodynamically stable (the maximum and minimum $p_{O_2}^\text{max}$ and $p_{O_2}^\text{min}$ are set by Ni oxidation and Al$_2$O$_3$ reduction - see~\hyperlink{linkSI}{SI}). 
For each sapphire termination, the Ni (111) termination may be  \textit{A}, \textit{B} or \textit{C}, corresponding to the classical description of the (111) plane stacking of FCC materials (i.e., \textit{...ABCABC...}) - these 3 terminations also represent Ni crystal shifts parallel to the interface (see~\hyperlink{linkSI}{SI}). 
Our DFT calculations show that the  termination C-Ni/sapphire is most stable for both 2Al and O sapphire terminations. 
For the O-terminated sapphire interface, the C-Ni termination corresponds to placing a Ni atom at the same position that would be occupied by an Al atom in perfect sapphire~\cite{zhang2002connection}. 
\begin{figure*}[!ht]
	\includegraphics[width=\textwidth]{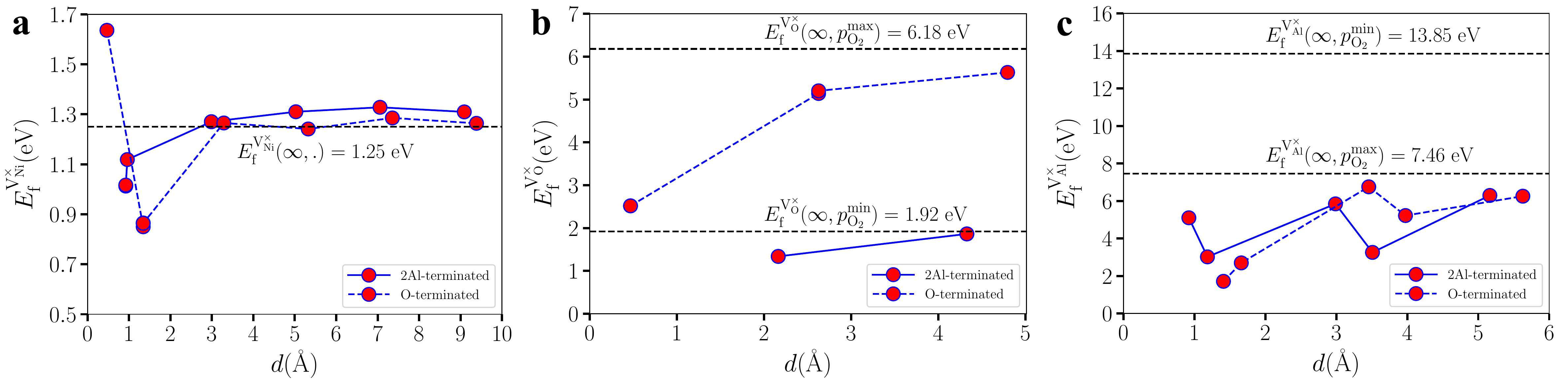}
	\caption{\textbf{Variation of vacancy formation energies ($E_\text{f}^{\text{V}_\text{Ni}^\times}$, $E_\text{f}^{\text{V}_\text{O}^\times}$, $E_\text{f}^{\text{V}_\text{Al}^\times}$) versus distance from the interface $\bm{d}$.} \textbf{a}, Ni vacancy formation energy for both O-terminated and 2Al-terminated sapphire interfaces. The horizontal dashed line represents the vacancy formation in bulk strained Ni (corresponding to M1 epitaxy). \textbf{b}, Neutral O vacancy formation energy for both the terminations. For the O-terminated case, the bulk O vacancy formation energy (dashed line) is for $p_{O_2}^\text{max}$ and  at $p_{O_2}^\text{min}$ for the 2Al-termination. \textbf{c}, Neutral Al vacancy formation energy for both the terminations. The bulk Al vacancy formation energies for the O-terminated and 2Al-terminated cases are for $p_{O_2}^\text{max}$  and $p_{O_2}^\text{min}$.}
	\label{fig:layer_energy}
\end{figure*}

Since diffusion in Ni is vacancy-controlled~\cite{barsoum2002fundamentals}, we determine vacancy formation energies in Ni and $\alpha$-Al$_2$O$_3$ as a function of distance $d$ from the Ni/sapphire interface and $p_{\text{O}_2}$; e.g., the formation energy of the neutral O-vacancy (i.e., formed by removing an O atom and all of its electrons) in Al$_2$O$_3$ is denoted by $E_{\text{f}}^{\text{V$_\text{O}^\times$}}(d,p_{\text{O}_2})$ in  Kr\"oger-Vink notation~\cite{kroger1956relations}. 
At the metal-ceramic interface, we expect the net charge on  point defects to be near zero since the Fermi level of the system will be pinned to that of the metal~\cite{yeo_metal-dielectric_2002}. Hence, unlike in bulk ceramics (see~\hyperlink{linkSI}{SI}), neutral point defects may be formed near both sides of the interface. 

We first calculate neutral vacancy formation energies in bulk Ni and $\alpha$-Al$_2$O$_3$ (\textit{d}=$\infty$); $E_\text{f}^{\text{V}_\text{Ni}^\times}(\infty,\cdot)$ ($p_{\text{O}_2}=$``$\cdot$'' indicates $p_{\text{O}_2}$-independence), $E_\text{f}^{\text{V}_\text{Al}^\times}(\infty,p_{\text{O}_2})$ and $E_\text{f}^{\text{V}_\text{O}^\times}(\infty,p_{\text{O}_2})$. 
$E_\text{f}^{\text{V}_\text{Ni}^\times}(\infty,\cdot)=1.51$ eV (unstrained, close to the experimentally found value of $1.6$ eV~\cite{wycisk1978quenching}) or $1.25$ eV (strained to M1 epitaxial relationship, -2.8\%). 
%This is close to the $1.6$ eV found experimentally (unstrained Ni)~\cite{wycisk1978quenching}. 
Similarly, since Schottky defects are more prevalent than Frenkel defects in $\alpha$-Al$_2$O$_3$~\cite{mohapatra1978dominant,lee2014identification}, we focus on Al and O vacancies on the sapphire side of the interface. 
For bulk sapphire, the neutral vacancy formation energies depend on oxygen partial pressure;  on oxygen sites  $E_\text{f}^{\text{V}_\text{O}^\times}(\infty,p_{O_2}^\text{min})=1.92$ eV and $E_\text{f}^{\text{V}_\text{O}^\times}(\infty,p_{O_2}^\text{max})=6.18$ eV. The neutral Al vacancy formation energy is $E_\text{f}^{\text{V}_\text{Al}^\times}(\infty,p_{O_2}^\text{min})=13.85$ eV  and $E_\text{f}^{\text{V}_\text{Al}^\times}(\infty,p_{O_2}^\text{max})=7.46$ eV (see \hyperlink{linkSI}{SI} Fig. 3). These vacancy formation energies suggest that near the interface the equilibrium vacancy concentration in Ni is much higher than either Al or O vacancies in sapphire.  

Figure~\ref{fig:layer_energy} shows the neutral Ni, Al, and O vacancy formation energies as a function of distance \textit{d} from the  O-terminated sapphire and  2Al-terminated sapphire/C-plane Ni interfaces. 
For the O-terminated sapphire interface, the Ni vacancy formation energy $E_{\text{f}}^{\text{V$_\text{Ni}^\times$}}(d,\cdot)$ drops from its bulk value $1.28$ eV far from the interface to $0.85$ eV (i.e., 66\% of the bulk value for pure Ni) two (111) Ni atomic planes from the interface. 
On the other hand, immediately adjacent to the sapphire in Fig.~\ref{fig:layer_energy}a, the Ni vacancy formation energy is very large, $1.64$ eV, likely because of the strong metal--oxygen bond.
While removing a Ni atom from this site is not as energetically costly as removing an Al atom from bulk sapphire (the Al vacancy formation energy is $7-13$ eV depending on $p_{\text{O}_2}$), it is higher than it would be for removing a Ni atom from a Ni crystal. 
The same trend also applies to the 2Al-terminated sapphire interface. 
Hence, the  Ni vacancy formation energy is much smaller near the interface than in the Ni interior for all Ni/$\alpha$-Al$_2$O$_3$ interfaces. 
This implies that the thermal concentration of Ni vacancies near the interface is much higher than elsewhere in Ni.

Figure~\ref{fig:layer_energy}b,c  shows that the O and Al vacancy 
formation energies  decrease from their bulk values as we approach the interface. 
For example, in the O-terminated sapphire case, the O vacancy formation energy decreases from 5.63 eV to 2.52 eV as it approaches the interface  (at $p_{\text{O}_2}^\text{max}$). 
Similarly, $E_\text{f}^{\text{V}_\text{Al}^\times}(d,p_{\text{O}_2}^\text{max})$ drops from 6.25 eV to 1.71 eV near the interface (the 2Al-terminated case is discussed in~\hyperlink{linkSI}{SI}). 

While the lowest formation energy point defect near the interface is the Ni vacancy $E_\text{f}^{\text{V}_\text{Ni}^\times}=0.85$ eV, the third lowest  is the Al vacancy in O-terminated sapphire $E_\text{f}^{\text{V}_\text{Al}^\times}=1.71$ eV (the second lowest energy is an oxygen vacancy). In this case, however,  the Al vacancy is replaced by a Ni interstitial and a vacancy on the Ni side of the interface (see Fig. 7 in \hyperlink{linkSI}{SI}). 
Hence consideration of defect complexes at the interface involving the low-formation energy defects (V$_\mathrm{Ni}$, V$_\mathrm{Al}$) reveals that the Ni vacancy concentration at the metal/ceramic interface is expected to be even higher than that suggested by the single point defect formation energies alone.

\vspace{-0.5cm}
\section*{I\lowercase{nterface transport}}
\vspace{-0.4cm}
\noindent
Vacancy defect-mediated diffusion is commonly characterized as $D = D_0 e^{-E_{\text{f}}^\text{V}/k_BT}e^{-E_{\text{m}}^\text{V}/k_BT}$, where the pre-exponential factor  $D_0$  accounts for crystal structure, the effective atomic vibration frequency, interatomic separation, correlation and entropy effects, $k_BT$ is the thermal energy, $E_{\text{f}}^\text{V}$ and $E_\text{m}^\text{V}$ are the vacancy formation and migration energies respectively~\cite{knauth2002solid}. 
The arrhenius terms describe the equilibrium vacancy concentration and  vacancy migration, respectively.
While in ceramics, the point defect density may be modified by doping, in metals it is usually dictated by equilibrium thermodynamics (vacancies are easily produced/annihilated by dislocation climb). 

Using nudged elastic band calculations~\cite{henkelman2000climbing}, we determine the barrier for Ni vacancy migration (see \hyperlink{linkSI}{SI}) parallel to the interface (second layer) to be 0.49 eV, which is $\sim1/2$ its bulk value.  
Since the  Ni vacancy formation energy at this location is $\sim0.7$ that in bulk Ni, our results are consistent with earlier discovered trends~\cite{angsten_elemental_2014} that showed that in elemental FCC and HCP metals both the vacancy formation  and  migration energies scale in the same manner (linearly)  with  cohesive energy (i.e., the drop in the vacancy energies is related to reduced cohesion at the interface compared with bulk Ni). 
Combining the vacancy formation and migration results reported here, these results suggest that at, for example, half the Ni melting point, the interface diffusivity should be $\ge 10^4$ times faster than in bulk Ni.
GB diffusivities in metals are typically $(10^4-10^6)$ faster than lattice diffusion~\cite{mehrer2007diffusion}, 
which implies that metal/ceramic interface diffusivity and GB diffusivities are comparable at the same homologous temperature.  
Therefore, these results demonstrate that Ni transport along the Ni/sapphire interface is extraordinarily fast (relative to bulk diffusion). 
This conclusion is valid for both the coherent interface case analyzed here in detail and the case where the interface is semicoherent (which should be faster). 

To explore the generalization of this result, we compare the formation energy of a vacancy at the metal/ceramic interface to that within the bulk metal in terms of the local bonding at these locations, 
as captured by  the metal/ceramic work of adhesion \textit{W}$_\text{ad}$ and the metal surface energy $\gamma_\text{m}$. 
We estimate the  ratio of the metal vacancy formation energy at the metal/ceramic interface to that in the bulk metal in terms of a simple, heuristic bond breaking model (see~\hyperlink{linkSI}{SI}) as
\begin{equation}
\frac{E_\text{f}^{\text{V}_\text{m}^\times}(0)}{E_\text{f}^{\text{V}_\text{m}^\times}(\infty)}= \frac{W_\text{ad}}{\gamma_\text{m}}. 
\label{ratio}
\end{equation}

\begin{figure}[bp]
	%\begin{figure*}[!ht]
	\centering
	\includegraphics[scale=0.64]{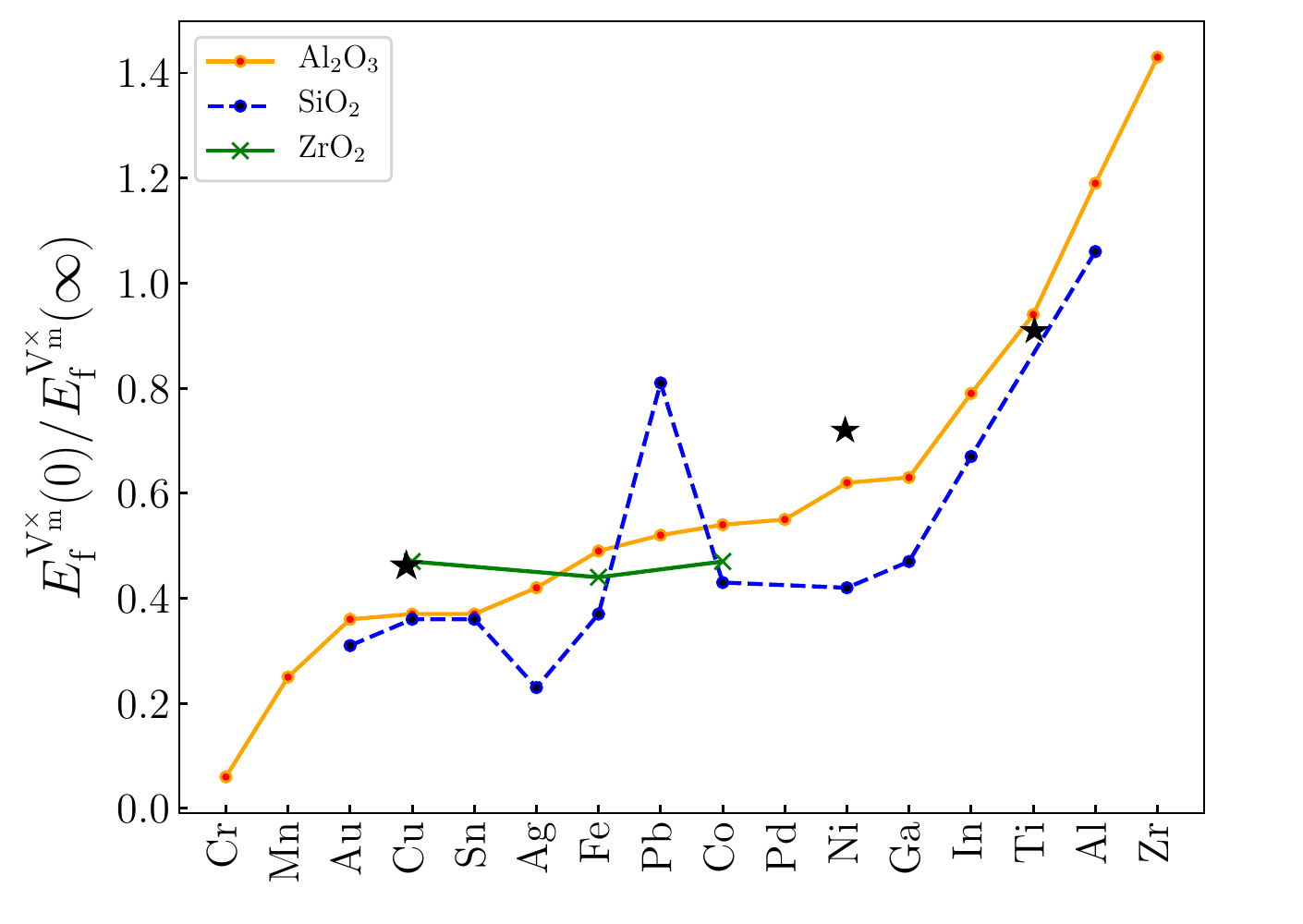}
	\caption{\textbf{The predicted ratio of the vacancy formation energies at the metal/ceramic interface to that in the bulk metal for several metal-ceramic systems according to equation~\eqref{ratio}}. The values for $\gamma_\text{m}$ and $W_\text{ad}$ are  from ~\onlinecite{tran2016surface,chatain1986adhesion,peden1991metal,campbell1997ultrathin,chatain1987estimation,sangiorgi1988wettability}. The metals are indicated along the horizontal axis and different ceramics by the three curves. The Ni/Al$_2$O$_3$, Cu/Al$_2$O$_3$ and Ti/Al$_2$O$_3$ data (black stars) are from direct DFT calculations. }
	\label{fig:ratio_trend}
\end{figure}

To test the applicability of this simple prediction, we compare it with the DFT results for  FCC-Ni/$\alpha$-Al$_2$O$_3$ (as described above), FCC-Cu/$\alpha$-Al$_2$O$_3$ and HCP-Ti/$\alpha$-Al$_2$O$_3$. 
As shown in Table~\ref{model} and Fig.~\ref{fig:ratio_trend}, the empirical descriptor ($W_\text{ad}/\gamma_\text{m}$) accurately predicts the ratio of $E_\text{f}^{\text{V}_\text{m}^\times}(0)/E_\text{f}^{\text{V}_\text{m}^\times}(\infty)$  to within 3$\%$ for Ti, 18\% for Cu, and 10\% for Ni. This is remarkable agreement given the simplicity of equation~\eqref{ratio}.
\begin{table}[!htbp]
	%\centering
	\caption{Vacancy formation energies of Ni, Cu and Ti at their interfaces with sapphire and in their bulk strained states, their ratio and the prediction as per  equation~\eqref{ratio} from~\onlinecite{tran2016surface,chatain1986adhesion,peden1991metal}.}
	\begin{tabular}{c|ccc|c|}
		%\cline{1-5}
		%& \multicolumn{3}{c}{DFT}  &  Descriptor \\%\multicolumn{2}{c}{Descriptor}\\
		&  & DFT &   &  Descriptor \\%\multicolumn{2}{c}{Descriptor}\\
		\cline{2-5}
		& \footnotesize{$E_\text{f}^{\text{V}_\text{M}^\times}(0)$ (eV)} & \footnotesize{$E_\text{f}^{\text{V}_\text{M}^\times}(\infty)$ (eV)} & \large{$\frac{E_\text{f}^{\text{V}_\text{m}^\times}(0)}{E_\text{f}^{\text{V}_\text{m}^\times}(\infty)}$}  &  \large{$\frac{W_\text{ad}}{\gamma_\text{m}}$}  \\
		%& \footnotesize{(eV)} & \footnotesize{(eV)} & & \\
		%Error $\%$\\
		%\midrule
		\cline{2-5}
		Ni   &  0.85 & 1.25 & 0.68 & 0.62 \\%&  10\\
		Cu  &  0.25 & 0.56 & 0.45 & 0.37 \\%& 19\\
		Ti  &  2.00 & 2.19 & 0.91 & 0.94 \\%& 3\\
		%\bottomrule 
	\end{tabular}
	\label{model}
\end{table}

We use this descriptor to predict the ratio of the vacancy formation energy at the metal/ceramic interface for a wide range of metal/ceramic systems based upon equation\eqref{ratio} and using data readily available  from the literature~\cite{tran2016surface,chatain1986adhesion,peden1991metal,campbell1997ultrathin,chatain1987estimation,sangiorgi1988wettability}. 
The resultant predictions are summarized in Fig.~\ref{fig:ratio_trend}. 
Given the correlation between activation energy for vacancy migration and vacancy formation,  we predict that metal diffusion along the interface  will be faster than bulk diffusion for systems in which $E_\text{f}^{\text{V}_\text{m}^\times}(0)/E_\text{f}^{\text{V}_\text{m}^\times}(\infty)<1$.  
Most of the metal/ceramic systems shown in Fig.~\ref{fig:ratio_trend} fall into this category, including Ni/Al$_2$O$_3$ (and Cu/Al$_2$O$_3$).  
For systems with $E_\text{f}^{\text{V}_\text{m}^\times}(0)/E_\text{f}^{\text{V}_\text{m}^\times}(\infty)\sim1$, diffusion in the bulk and at the interface will be comparable (Ti/Al$_2$O$_3$) and for others ($E_\text{f}^{\text{V}_\text{m}^\times}(0)/E_\text{f}^{\text{V}_\text{m}^\times}(\infty)>1$) interface diffusion is slower than in the bulk.
The case of Ti/Al$_2$O$_3$, for which we have DFT data, is near the cusp - the interface diffusivity should be comparable with bulk diffusion in Ti.
Amongst the cases shown in Fig.~\ref{fig:ratio_trend} are many metal/ceramic pairs that are commonly used across a wide range of technologies.
We note that the \{Cr, Mn, Au, Cu, Sn\}/Al$_2$O$_3$ and that \{Au, Cu, Sn, Ag, Fe, Co Pd, Ga\}/SiO$_2$ interfaces all show $E_\text{f}^{\text{V}_\text{m}^\times}(0)/E_\text{f}^{\text{V}_\text{m}^\times}(\infty)<0.5$, suggesting that all of these interfaces will exhibit extremely fast metal transport along these interfaces. 
However, given that extremely fast metal atom diffusion along metal/ceramic interfaces is the rule rather than the exception, fast metal/ceramic interface diffusion should not be considered anomalous after all.

We demonstrated both experimentally and computationally that diffusion at the Ni/sapphire is surprisingly fast. 
Our first-principles vacancy formation and migration energy calculations demonstrate that this is a result of relatively low cohesion at this interface compared with bulk Ni.
This observation suggests a simple descriptor for diffusion at the interface compared with the bulk based upon readily available experimental and/or first-principles results.
Based on this descriptor, we conclude that for most metal/ceramic systems (we examined close-packed metals and sapphire, silica, zirconia), interface diffusion is fast compared with the bulk and comparable with metal grain boundary diffusivities in many cases; yet there are exceptions (as determined based on interface cohesion).
Systems where the metal only weakly wets (or does not wet) the ceramic, the interface diffusivity will be high; inversely, where the tendency for wetting is strong, the interface diffusivity will be low (all relative to the bulk metal). 
This suggests that alloying to modify wettability also affects interface diffusion kinetics. 
This simple result provides easily applicable guidance for material design in a wide range of applications; especially in the energy and microelectronics industries. 

\subsection*{Methods}
\noindent
Methods are described in details along with statements of data availability in the \href{url}{online version of the paper}.

% Bibliography
\bibliographystyle{unsrt}
\bibliography{citations}
%\printbibliography 

\subsection*{Acknowledgements}
\noindent
The authors grateful acknowledge the the support of the (AK and DJS) US National Science Foundation grant 1609267, (HB and ER) US-Israel Binational Science Foundation grant 2015680 and Israel Science Foundation grant 1628/15. Work partly performed under the auspices of US DOE by LLNL under Contract DE-AC52-07NA27344. The authors acknowledge fruitful discussions with Jian Han, Jianwei Sun, Anuj Goyal, Dor Amram, Bilge Yildiz and Mostafa Youssef, and Christoff Freysoldt.
\subsection*{Author contributions}
\noindent
A.K., E.R., D.J.S. designed the research; A.K.,H.B. performed the computational and experimental research respectively; L.K., E.R. developed the diffusion model; L.K. performed the numerical calculations, M.W.F., V.L. contributed to discussions and analysis; the diffusion descriptor was proposed by A.K, E.R. and D.J.S; and A.K., H.B., E.R., D.J.S. wrote the paper.

\subsection*{Additional information}
\noindent
Supplementary information is available in the \href{url}{online version} of the paper.
\subsection*{Conflict of interest}
\noindent
The authors declare no conflict of interest.
\end{document}